\PassOptionsToPackage{svgnames}{xcolor}
\documentclass[12pt]{article}

\usepackage{needspace,amsmath,empheq,amssymb,amsfonts,amsthm,setspace,tikz,dsfont,wrapfig,lipsum,natbib, array,colortbl,mathrsfs,pifont,wasysym,nicefrac,multicol,booktabs,makecell,multirow,thm-restate,caption, subcaption,setspace,makecell}
\usetikzlibrary{patterns}
\usepackage{etoolbox}
\usepackage{thmtools,thm-restate}
\usepackage{caption}
\usepackage[all]{nowidow}
\usepackage[makeroom]{cancel}

\def\equationautorefname~#1\null{%
	Expression~(#1)\null
}

\usepackage[a4paper]{geometry}
\usepackage{hyperref}

\hypersetup{                                             
    pdffitwindow=true,                                  
    pdfstartview={XYZ null null 1.00},                   
    pdfnewwindow=true,                                   
    colorlinks=true,                                     
    linkcolor=DarkBlue,                                     
    citecolor=DarkBlue,                                     
    urlcolor=DarkBlue,                                      
    pdfauthor = {Gabriel Ziegler},
    pdfkeywords = {Informational Robustness, Common Belief of Rationality},
    pdftitle = {IRCBR},
    pdfsubject = {IRCBR},
    pdfpagemode = UseNone}

  \usepackage{pgfplots}
  \pgfplotsset{compat=newest}
  \usetikzlibrary{plotmarks}
  \usetikzlibrary{arrows.meta}
  \usepgfplotslibrary{patchplots}
  \usepackage{grffile}
  \usetikzlibrary{positioning}

\pgfplotsset{plot coordinates/math parser=false}
\newlength\figureheight
\newlength\figurewidth
\usepgfplotslibrary{colormaps}
\usetikzlibrary{pgfplots.colormaps} 

\usepackage{fancyhdr,ifthen}
\fancypagestyle{appendix}{
\fancyhead{}
\fancyhead[R]{\ifthenelse{\isodd{\value{page}}}{\thepage}{}}
\fancyhead[L]{\ifthenelse{\isodd{\value{page}}}{\textsc{}}{\thepage}}

\fancyfoot{}
}

\newtheorem{assumption}{Assumption}
\newtheorem{lemma}{Lemma}
\newtheorem{proposition}{Proposition}

\newtheorem{corollary}{Corollary}

\theoremstyle{definition}

\usepackage{titlesec}

\titleformat*{\section}{\large\scshape\centering}
\titleformat*{\subsection}{\scshape\centering}
\titleformat*{\subsubsection}{\itshape}
\titleformat*{\paragraph}{\large\bfseries\centering}
\titleformat*{\subparagraph}{\large\bfseries\centering}

\DeclareMathOperator{\ppv}{PPV}
\DeclareMathOperator{\npv}{NPV}

\makeatletter
\DeclareRobustCommand\citepos
  {\begingroup
   \let\NAT@nmfmt\NAT@posfmt
   \NAT@swafalse\let\NAT@ctype\z@\NAT@partrue
   \@ifstar{\NAT@fulltrue\NAT@citetp}{\NAT@fullfalse\NAT@citetp}}

\let\NAT@orig@nmfmt\NAT@nmfmt
\def\NAT@posfmt#1{\NAT@orig@nmfmt{#1's}}

\makeatother

\title{\vspace{-1em}Binary Classification Tests, Imperfect Standards, and Ambiguous Information\footnote{Thanks to Dan Sacks, Charles Manski, and J\"{o}rg Stoye for literature pointers. I thank Filip Obradovic for very valuable comments.  Christopher Stapenhurst provided excellent research assistance. All errors are of course mine.}}

\author{
Gabriel Ziegler\footnote{The University of Edinburgh, School of Economics; 31 Buccleuch Place, Edinburgh, EH8 9JT, UK; \href{mailto:ziegler@ed.ac.uk}{\tt ziegler@ed.ac.uk}.}
} 

\begin{document}

\maketitle
\thispagestyle{empty}
\vspace*{-2em}
\begin{abstract}
\small
New binary classification tests are often evaluated relative to a pre-established test. For example, rapid Antigen tests for the detection of SARS-CoV-2 are assessed relative to more established PCR tests. In this paper, I argue that the new test can be described as producing ambiguous information when the pre-established is imperfect. This allows for a phenomenon called dilation---an extreme form of non-informativeness. As an example, I present hypothetical test data satisfying the WHO's minimum quality requirement for rapid Antigen tests which leads to dilation. The ambiguity in the information arises from a missing data problem due to imperfection of the established test: the joint distribution of true infection and test results is not observed. Using results from Copula theory, I construct the (usually non-singleton) set of all these possible joint distributions, which allows me to assess the new test's informativeness. This analysis leads to a simple sufficient condition to make sure that a new test is not a dilation. I illustrate my approach with applications to data from three COVID-19 related tests. Two rapid Antigen tests satisfy my sufficient condition easily and are therefore informative. However, less accurate procedures, like chest CT scans, may exhibit dilation.

\vspace{0.2cm}

\end{abstract}
\cleardoublepage
\setcounter{page}{1}
\setstretch{1.3}
\section{Introduction}
\label{section:introduction}
An important aspect of evaluating a new diagnostic test is to assess its accuracy. Intuitively, a sensible binary test should have test results highly correlated with the underlying health condition. In other words, a positive test result should be likely if and only the tested person is indeed infected or sick.\footnote{In the following, I will not differentiate between being infected and being sick.} However, establishing whether a person is truly infected is often costly or even impossible. Therefore, a new test is analyzed relative to an established test. An established test is perfect when a positive test result occurs if and only if the person is truly infected. The medical literature calls these perfect tests a ``gold standard'' \citep{watson-etal-2020}. In these situations the joint distribution of the new test's outcomes and the underlying true health condition is the \emph{same} as the joint distribution of test results from both tests. Thus, this observed joint distribution can be used to evaluate the new test's accuracy.

In practice, however, a perfect reference test does not exist. In such a case, the researcher would need the joint distribution of the health conditions and the outcomes of both tests.\footnote{Depending on the question, it might suffice to consider the joint distribution of the new test's outcomes and the underlying true health condition. For example, the analysis \autoref{subsection:newtest} requires only this bivariate joint distribution. The trivariate distribution is needed to evaluate the informativeness of performing both tests as discussed in \autoref{subsection:additest}.} This overall joint distribution is not observable (or maybe only if the researcher incurs a high costs for obtaining the data). This missing data problem leads to two distinct problems: ($i$) the marginal distribution of the underlying health condition is missing and ($ii$) the \emph{correlation} between new test's outcome and health status is missing too.\footnote{In the introduction, I use the word correlation loosely and informal.} The latter of these problems will introduce ambiguity in the information provided by the new test.

The first of these problems, missing data about the underlying health condition, is well-known. Recently, \cite{manski-molinari-2021} use methods known from the literature on partial identification to provide bounds on prevalence---the fraction of infected people in the population.\footnote{Similar approaches were used by \cite{stoye-2020} and \cite{sacks-etal-2020}.}  Measuring prevalence is different from the usual inference problem because the tested population might not be representative of the overall population. Here the data are observed selectively which corresponds to a \emph{selection problem} as introduced by \cite{manski-1989}. Furthermore, \cite{manski-2020} illustrates how this problem carries over to evaluating accuracy of new tests in the context of COVID-19 Antibody tests maintaining the assumption of a perfect reference test.

The second problem of missing data about the correlation is different in nature and avoided when a perfect reference is available. Even if one would assume knowledge of prevalence,  potentially multiple 'correlation structures' are consistent with the observed data. The reason for this multiplicity is well-known from copulas as studied in probability theory. Knowledge of prevalence provides the marginal distribution of the health condition, whereas the observed testing data provides (a bivariate) marginal distribution. In general, there are multiple (trivariate) joint distributions with these marginal distributions. Due to this multiplicity a simple, unambiguous interpretation of the new test is not possible. Without knowledge of prevalence, the problem identified before carries over and therefore exacerbates the overall multiplicity. However, as discussed in more detail later, the ambiguous information stems only from the missing data on correlation and therefore occurs whether or not the researcher has knowledge about prevalence.

In this paper, I provide a theoretic framework that combines insights from \cite{manski-molinari-2021} and \cite{stoye-2020} about selective testing with the missing correlation data due to an imperfect reference test. Within this framework, it is possible to address informativeness of both tests. First, \autoref{proposition:npv_y} shows that the established test's negative predictive value\footnote{A test's negative predictive values is the probability of being healthy conditional on obtaining a negative test result. Another important informativeness measure is the positive predictive value, which is the probability of being infected conditional on a positive test result. I will assume throughout that the established has a perfect predictive value in line with the application to SARS-CoV-2 testing.}   is usually not given by a unique number, but it always informative nevertheless. This multiplicity arises because of problem ($i$) only. Then, I analyze the new test's informativeness for the test population only. The focus on the tested population simplifies the algebra and furthermore shuts down the ambiguity about prevalence (cf.\ problem ($i$)) and therefore allows me to study the essence of ambiguous information for the new test in separation (problem ($ii$) only). Finally, I study the implications on informativeness if both effects are present.

Studying the informativeness of tests has a long tradition in probability theory, statistics, economics, and philosophy. \cite{blackwell-1951,blackwell-1953} introduces a notion of ``(more) informative'' for (what is now called Blackwell) experiments.\footnote{\cite{deoliveira-2018}  provides a more recent treatment.} An experiment is a mapping from states of the world to a distribution over signals. In the current setting, an experiment is a function that associates a distribution over test results to each of the possible health conditions, i.e.\ for being infected and for being healthy. In such a setting, the value of information is defined as the amount a Bayesian decision maker is willing to pay for the experiment. Since every experiment is more informative than an uninformative experiment,\footnote{An experiment is uninformative if the mapping mentioned above is a constant function?} \citeauthor{blackwell-1951}'s theorem shows that the value of information is (weakly) positive for every Bayesian decision maker.\footnote{More generally, \citeauthor{blackwell-1951} characterizes his notion of ``more informative'' with the requirement that every Bayesian decision maker has a higher value of information for the more informative experiment.}  Ideally a diagnostic test should satisfy \citeauthor{blackwell-1951}'s definition of an experiment in order to ensure that it is always informative. However, this is typically only true for the established test in my framework.

The new test fails to be a Blackwell experiment because it does not map each state to a unique distribution over test results. Rather, due to the multiplicity of joint distributions, there is a \emph{set} of distributions over test results for a given health condition.\footnote{Formally, the new test can be seen a correspondence or set-valued function.} Therefore, \citeauthor{blackwell-1951}'s informativeness notion does not apply to the new test. Furthermore, the value of information needs to be adjusted because a Bayesian analysis does not readily apply with sets of probabilities. Such a situation is usually referred to as a situation of ``ambiguity'' and the literature has identified several extensions of Bayesian decision making to the realm of ambiguity.\footnote{\cite{machina-siniscalchi-2014} provide a recent overview about this topic.}

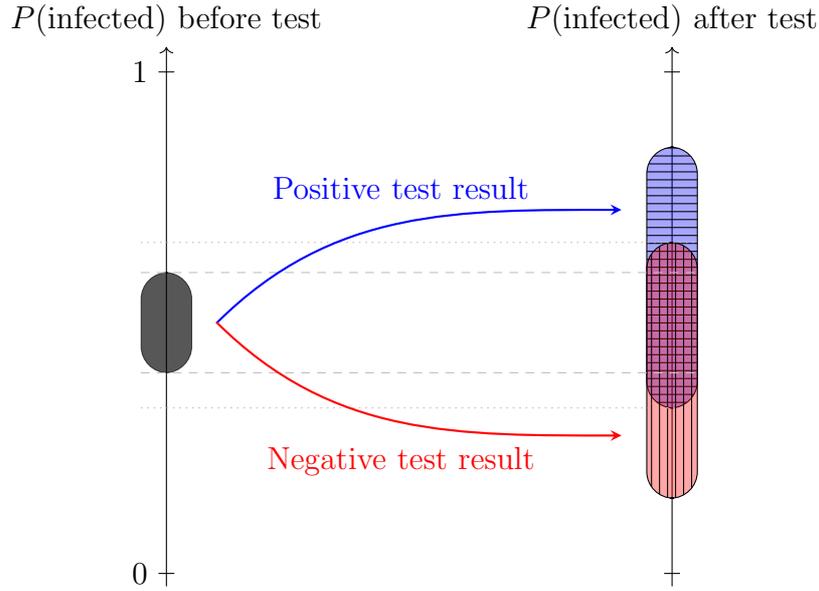
\begin{figure}[!ht!]\centering
	\begin{tikzpicture}[line cap=round, line join=round,
		x=5cm, y=5.0cm, scale=1.33,
		dot/.style={circle,fill=black,minimum size=4pt,inner sep=0pt,
			outer sep=-1pt}]
		\draw [->,color=black] (0,-0.025) -- (0,1.05) node [above] {$P(\text{infected})$ before test};
		\draw [->,color=black] (1,-0.025) -- (1,1.05) node [above] {$P(\text{infected})$ after test};
		\draw [-,color=black] (0.015,1) -- (-0.015,1) node [left] {$1$};
		\draw [-,color=black] (1.015,1) -- (.985,1);
		\draw [-,color=black] (0.015,0) -- (-0.015,0) node [left] {$0$};
		\draw [-,color=black] (1.015,0) -- (.985,0);
		\draw[rounded corners=10pt, black, fill=black, opacity=0.66] (-0.05,0.4) rectangle (0.05,0.6);
		\draw[preaction={fill, blue, opacity=0.35}, rounded corners=10pt, black,pattern=horizontal lines,pattern color=black]
		(0.95,0.33) rectangle (1.05,0.85);
		\draw[preaction={fill, red, opacity=0.35}, rounded corners=10pt, black,pattern=vertical lines,pattern color=black]
		(0.95,0.66) rectangle (1.05,0.15);
		\draw [dashed,color=black!25] (-0.05,0.4) -- (1.05,0.4);
		\draw [dashed,color=black!25] (-0.05,0.6) -- (1.05,0.6);
		\draw [dotted,color=black!25] (-0.05,0.66) -- (1.05,0.66);
		\draw [dotted,color=black!25] (-0.05,0.33) -- (1.05,0.33);
		\draw [->,>=stealth,out=45,in=180,color=blue, thick] (0.1,0.5) to node [yshift=5, midway, above, blue] {Positive test result} (0.9,0.725);
		\draw [->,>=stealth,out=-45,in=180,color=red, thick] (0.1,0.5) to node [yshift=-5, midway, below, red] {Negative test result} (0.9,0.275);
	\end{tikzpicture}
	\caption{A diagnostic test as dilation} \label{fig:dilation_illustration}
\end{figure}

Instead of defining the value of information for a specific decision criterion in such a situation, I adopt a very weak notion of informativeness: the diagnostic test is informative if and only if it is not a \emph{dilation}. \cite{seidenfeld-wasserman-1993} introduce the notation of dilation for situations with sets of probabilities. In the current context, a dilation occurs if, no matter what test result is obtained, the set of probabilities conditional on this information contains the original set of probabilities. \autoref{fig:dilation_illustration} illustrates an example of a dilation. Here, the set of probabilities indicating the infection likelihood before the test (black set) lies within both sets after the test result (blue for a positive result and red indicating the set after a negative result). Thus, in a sense, the decision maker is worse-off after taking the test than before taking the test \emph{no matter} what the test result is. For this reason, \citeauthor{seidenfeld-wasserman-1993}  call a dilation a ``counterintuitive phenomenon'' and \cite{gul-pesendorfer-2018} refer to it as ``all news is bad news''.

My framework allows to fully characterize when a new diagnostic test is a dilation (cf.\ \autoref{eq:dilation}). Since the definition of informativeness for tests is rather weak, any reasonable test should satisfy this criterion. The characterization provides a method to verify whether the new test is informative.

The \cite{who-2020} recommends a minimum standard of accuracy for rapid Antigen tests.\footnote{For the informed reader, the \citeauthor{who-2020} recommends a sensitivity of at least $80\%$ and a specificity of at least $97\%$. These measures will be formally introduced and defined later.} Usually a PCR test is the established test used to evaluate these Antigen tests \citep{esbin-etal-2020}. \autoref{tab:dilation_test} illustrates hypothetical test data, which fulfill the minimal requirements of the \citeauthor{who-2020}. However, as the analysis will reveal, this test is actually a dilation and therefore not informative.\footnote{This statement depends on the accuracy of the established PCR test. A dilation occurs only if the PCR has sensitivity at the lower range identified by the literature.}
\begin{restatable}{table}{dilationtest}
	\centering
	\caption{Dilation Test Data}
	\label{tab:dilation_test}
	\begin{tabular}{@{}cccc@{}}
		\toprule
		$z \backslash y$ & $y=0$ & $y=1$ & Sum	\\ \midrule
		$z=0$ & $39.5\%$ & $11.5\%$ & $51\%$\\
		$z=1$ & $1.2\%$ & $47.8\%$ & $49\%$\\ \midrule
		Sum & $40.7\%$ & $59.3\%$ \\
		\bottomrule 
		\multicolumn{4}{c}{\parbox{5cm}{\vspace*{0.25em} \centering \scriptsize $y=1$ indicates a positive PCR-test result (i.e.\ the established test). $z=1$ denotes a positive Antigen test.}}
	\end{tabular}
\end{restatable}
For minimum required accuracy standards, the dilation characterization provides an easy to verify sufficient condition to avoid dilation. The new test is informative (in the population of tested people) if
\begin{empheq}[box=\fbox]{align*}
	\text{Fraction of people with}&\text{ a positive new test } 
	\\ &\leq  \\
 	\text{minimum sensitivity of new test } &\times \text{ sensitivity of established test}.
\end{empheq}

Besides theoretic applications of dilation, there is not much empirical evidence in the literature yet. Recently, economists started to investigate dilation and ambiguous information experimentally. The only experiment focusing on how decision makers react to dilation and relate the behavior to the value of information is conduced by \cite{shishkin-ortoleva-2020}. To the best of my knowledge, the possible occurrence of dilation with diagnostic tests (or SARS-CoV-2 tests more specifically) is the first observation of this phenomenon `in the field.'\footnote{\cite{manski-2018} mentions that a dilation might occur in a different medical context, but does not address this issue further.}

Of course, researchers studying diagnostic tests are well aware of the general issues addressed here. The problem of selection leading to unobserved prevalence is known as \emph{Verification Bias}, whereas the problem arising from unobserved correlation due to an imperfect reference test is descriptively named \emph{Imperfect Gold Standard Bias}. \cite[Chapters 10--11]{zhou-etal-2014} This paper is not the first to document that either of of these problem leads to non-identified models; rather the novelty of this paper comes in the approach. Diagnostic test research seeks to avoid non-identified models by introducing additional assumptions and then address resulting biases relative to a baseline assumption. Proposed methods include simply imputing missing data or considering more sophisticated correction methods. Moreover, the two problems are often addressed separately. By contrast, my framework requires minimal assumptions and addresses both problems  simultaneously.\footnote{\cite{reitsma-etal-2009} provide a flowchart as guidance for applied researchers to address several problems arising when  establishing accuracy of diagnostic tests. The two problems addressed here are in two distinct branches of the flowchart.}  

\section{Main Analysis} \label{section:main}
I consider the following situation. Let $x=1$ denote that a person is infected and $x=0$ if the person is healthy. Initially, there is binary test available where $y=1$ indicates a positive test result and $y=0$ a negative result. Finally, a new test is introduced which again can be either positive ($z=1$) or negative ($z=0$).

Let $P(x,y,z)$ denote the population distribution under consideration with $p:=P(x=1)$ denoting prevalence. However, the population distribution is not directly observable. This is, of course, almost always the case, because a researcher usually only observes a sample from the population distribution. This leads to the usual inference problem. Throughout, I will abstract  away  from inference altogether.  Instead, the data is given for people who were tested to obtain data on the new test. For this denote tested people with $t=1$ and $t=0$ otherwise. Then, the data are given by $P(y,z|t=1)$ and I assume that $P(t=1)>0$.\footnote{Furthermore, the following logical implications of (not) being tested hold: ($i$) $t=0 \implies z=0$ and ($ii$) $ z=1 \implies t=1$. Note that $y=1$ is possible even if not tested, because the participation pool concerns only the new test.}\textsuperscript{,}\footnote{Equivalently, the data is given by sensitivity and specificity of the new test \emph{relative} to the established test with the additional information about how many established or new tests had a positive result.}

Furthermore, since the established test is well-known, precise information about the sensitivity and specificity of this test is available as well. The following assumption ensures that both of these measures are well defined.
\begin{assumption}[Non-trivial prevalence]\label{assumption:non-trivial}
	The population satisfies $p\in(0,1)$.
\end{assumption}

With this assumption, sensitivity and specificity for the initial test are respectively defined as:
\begin{align}
	P(y=1|x=1) &= \frac{P(x=1,y=1)}{p} =: \sigma, \; \text{ and} \\
	P(y=0|x=0) &= \frac{P(x=0,y=0)}{1-p}.
\end{align}

As discussed in \cite{manski-2020}, for decision making sensitivity and specificity are not the relevant measures. The relevant measures are \emph{positive predictive value (PPV)} and \emph{negative predictive value (NPV)}. For the established test these measures can be obtained from specificity and sensitivity via Bayes' rule if prevalence $p$ and $P(y=1)$ are known:
\begin{align*}
	\ppv_y &:= P(x=1|y=1) = \frac{p}{P(y=1)}P(y=1|x=1) = \frac{p}{P(y=1)}\sigma\\
	\npv_y &:= P(x=0|y=0) = \frac{1-p}{P(y=0)}P(y=0|x=0).
\end{align*}
Since the tested people are usually not representative of the overall population,\footnote{For example, supposedly infected people may be oversampled in order to get meaningful results.\label{footnote:oversampling}} even for the established test these two measures are not point-identified. \citep{manski-molinari-2021,manski-2020,stoye-2020} 

To simplify the analysis and in-line with the application to SARS-CoV-2 testing, I also consider the following three baseline assumptions.
\begin{assumption}[No False-Positives for established test] \label{assumption:no_false-pos}
	The population satisfies $P(x=0, y=1)=0$.
\end{assumption}
\autoref{assumption:no_false-pos} implies that the established test achieves a maximum specificity and $\ppv_y$ of $1$.\footnote{This holds because $
		P(x=0,y=0) = P(x=0,y=0) + P(x=0, y=1) = P(x=0) = 1-p$ and 
$P(x=1, y=1) = P(x=1, y=1) + P(x=0, y=1) = P(y=1)$.}

Additionally, I will assume \emph{test-monotonicity} as in \cite{manski-molinari-2021}, meaning conditional on being tested the probability of being infected is greater than if not being tested.\footnote{This might not be true, if there is voluntary enrollment into the testing pool. However, for establishing the accuracy of new tests this assumptions seems to be applicable often. See \autoref{footnote:oversampling}.} 
\begin{assumption}[Test-monotonicity] \label{assumption:test_monotone}
	The population satisfies $P(x=1|t=1)\geq P(x=1|t=0)$.
\end{assumption}

Lastly, I assume that the established test's sensitivity does depend on the underlying health status $x$, but not on whether the person is in the testing pool $t=1$.\footnote{Recall that the testing pool is obtained for the new test. This assumption might be violated, if, for example, the medical staff performing the established test for the testing pool is extra careful. In this case, the established test might be more sensitive for the testing pool.}
\begin{assumption}[Health-sufficiency] \label{assumption:health-suff}
	The population satisfies $P(y=1|x=1,t=1) = P(y=1|x=1) = \sigma$.
\end{assumption}

To reduce cumbersome lengthy notation in the following, I will use this simplified notation henceforth:
\begin{align*}
	\gamma &:= P(y=1|t=1) &\ldots &\text{ test yield for established test} \\
	\zeta &:= P(z=1|t=1) &\ldots &\text{ test yield for new test} \\
	\tau &:= P(t=1) &\ldots &\text{ measure of data representativeness}.
\end{align*}
To avoid trivial cases, assume that $\gamma, \zeta, \tau >0$. Note that $\tau$ has a slightly different interpretation as in \cite{manski-molinari-2021} or \cite{stoye-2020}. Here, $\tau=1$ means the data $P(y,z|t=1)$ is perfectly representative of the overall population. In particular, such a parameter value implies no oversampling of infected participants.\footnote{See \autoref{footnote:oversampling} for why such an assumption might be problematic.} In particular, even if the participation pool is small (as is often the case), this does not mean that $\tau$ should be close to zero.\footnote{A small participation pool might worsen the statistical inference problem: suppose the participation pool is perfectly representative but small. In this case $\tau=1$, but inference usually relies on some sort of central limit theorem which would not be appropriate in this scenario. However, recall that I abstract away from inference problems as mentioned above.} 

With this notation, we have $P(z=1)=\tau \zeta$ since the new test is positive only if the person was tested. Furthermore, \autoref{assumption:no_false-pos} combined with \autoref{assumption:health-suff} gives  $P(x=1|t=1) = \nicefrac{\gamma}{\sigma}$. Then, the Law of Total Probability together with \autoref{assumption:test_monotone} provides sharp bounds\footnote{A bound for a given set is called \emph{sharp} if the bound itself is a member of this set.} on prevalence $p \in \left[ \tau \nicefrac{\gamma}{\sigma}, \nicefrac{\gamma}{\sigma} \right]=:\left[\underline{\chi}, \overline{\chi}\right]$ because

\begin{align*}
	p := P(x=1) = \underbrace{P(x=1|t=1)}_{=\frac{\gamma}{\sigma}} \tau + \underbrace{P(x=1|t=0)}_{\in \left[0, \frac{\gamma}{\sigma}\right] \text{ by \autoref{assumption:test_monotone}}}(1-\tau).
\end{align*}
In turn, bounds on the established test's overall positivity rate are implied by sensitivity $\sigma$ and \autoref{assumption:no_false-pos}: $P(y=1) = p\sigma \in \left[\tau \gamma, \gamma\right]$.

Since we consider the non-trivial case of $p\in(0,1)$, consistency of the data with the maintained assumptions requires the established test's sensitivity to be sufficiency high , i.e.\ $\gamma < \sigma \leq 1$. In turn, the assumptions imply $P(y=1) \in (0,1)$.

\subsection{The established test} \label{subsection:establishedtest}
\autoref{assumption:no_false-pos} implies a perfect positive predictive value for the established test ($\ppv_y=1$). However, the negative predictive value is only partially identified and \autoref{proposition:npv_y} provides sharp bounds.
\begin{proposition} \label{proposition:npv_y}
	Under \autoref{assumption:non-trivial}--\autoref{assumption:health-suff}, the established test's negative predictive value is sharply bounded as follows:
	\begin{align*}
		\npv_y \in \left[\frac{1}{\sigma}\frac{\sigma - \gamma}{1-\gamma} , \frac{1}{\sigma}\frac{\sigma - \tau\gamma}{1-\tau\gamma} \right].
	\end{align*}
\end{proposition}

\begin{proof}
	Fix $\alpha = P(y=1) \in \left[\tau \gamma, \gamma\right]$ and define prevalence as a function of $\alpha$ by $P_{\alpha}(x=1)=\alpha/\sigma$. Then
	\begin{align*}
		\npv_y(\alpha) = \frac{1-P_{\alpha}(x=1)}{1-\alpha} \underbrace{P(y=0|x=0)}_{=1\text{ by \autoref{assumption:no_false-pos}}}= \frac{1}{\sigma}\frac{\sigma - \alpha}{1-\alpha}
	\end{align*}
	Since $\sigma - \alpha \leq 1 - \alpha$ for all $\alpha \in \left[\tau \gamma, \gamma\right]$, $NPV_y(\cdot)$ is decreasing. Therefore, $\npv_y \in \left[\npv_y(\gamma), \npv_y(\tau \gamma)\right]$.
\end{proof}	

With this in hand, the established test's informativeness can be analyzed. \autoref{tab:info_pcr} summarizes the prevalence before and after observing a test result from the established test, which are the relevant measures for defining informativeness (cf.\ \autoref{fig:dilation_illustration}). Formally, the established test is a \emph{dilation} if every possible prevalence $p \in [\underline \chi, \overline \chi]$ is a possible value of both $P(x=1|y=1)=\ppv_y$ and  $P(x=1|y=0)=1-\npv_y$. Obviously, a positive test result gives perfect knowledge due to the maintained assumptions. Thus, the established test cannot be a dilation. On the other hand, a negative result lowers the lower and upper bound of prevalence conditional on a negative result because $\tau \gamma \leq \gamma < \sigma$.\footnote{Of course, this has to hold since the Law of Total Probability holds pointwise.} Furthermore, the interval width for any test result shrinks the set of possible values for prevalence conditional on either test result.\footnote{It is obvious for a positive test result. For a negative result, note that the width strictly increases if and only if $1-\sigma > (1- \gamma)(1-\tau\gamma)$, which is equivalent to $\tau(1+\gamma) > \frac{\sigma}{\gamma}+1 \geq 2$ leading to a contradiction.} In this sense, the established test is not just informative (i.e. not a dilation), but also strictly shrinks the size of the set of possible prevalence values after a negative test result.
\begin{table}[ht!]
	\centering
	\caption{Informativeness of the established test}
	\label{tab:info_pcr}
	\begin{tabular}{@{}c|ccc@{}}
		\toprule
		$P(x=1|\cdot)$ & Lower bound & Upper bound & Interval Width	\\ \midrule
		Prior testing & $\tau\frac{\gamma}{\sigma}$ & $\frac{\gamma}{\sigma}$ & $\frac{\gamma}{\sigma}\left(1-\tau\right)$\\[0.5ex]
		Positive result ($y=1$) & $1$ & $1$ & $0$\\[0.5ex]
		Negative result ($y=0$) & $\frac{\tau\gamma}{\sigma}\frac{1-\sigma}{1-\tau\gamma}$ & $\frac{\gamma}{\sigma}\frac{1-\sigma}{1-\gamma}$ & $\frac{\gamma}{\sigma}\frac{1-\tau}{1-\tau \gamma}\frac{1-\sigma}{1-\gamma}$\\
		\bottomrule
		\multicolumn{4}{c}{\parbox{12cm}{\vspace*{0.25em} \centering \scriptsize Remark: The second row corresponds to $\ppv_y$ and the third row is given by $1-\npv_y$.}}
	\end{tabular}
\end{table}

It is well known that knowledge of prevalence is needed in order to apply Bayes' rule to obtain NPV. Since in most applications prevalence is not known, a common practice is to assume a given prevalence level. For example the United States Food and Drug Administration \citep{fda-2020} assumes a prevalence of $5\%$ to calculate PPV and NPV.  If such an assumption ($p=\chi$) is added to the maintained assumptions, then $P(y=1)=\chi \sigma$ and furthermore $P(y=1|t=0) = \frac{ \chi \sigma - \gamma \tau}{1-\tau}$.\footnote{Alternatively, one could drop the assumption that $P(t=1)=\tau$ is known exactly. In this case (and allowing $P(y=1|t=0) \in [0, \gamma]$ as in the general case) the assumed prevalence bounds $\tau$. Calculations show that $\tau \in  \left[0, \frac{\chi \sigma}{\gamma}\right]$. Since the lower bound is always $\tau_{\min}=0$, we do not find this case very interesting.} This additional assumption allows to exactly pin down the established test's NPV as $\frac{1-\chi}{1-\chi\sigma}$ and therefore $P(x=1|y=0) = \chi \frac{1-\sigma}{1-\chi \sigma}$. Thus, this additional assumption not only assumes away the ambiguity about prevalence, but also illustrates that the established test does not provide ambiguous information itself.\footnote{Technically, the established test is an experiment \'{a} la \cite{blackwell-1951}, where sensitivity and specificity can be seen as functions mapping (health) states to distributions over signals (i.e. test results). As mentioned in the introduction, this implies that the established test's value of information is (weakly) positive under these assumptions.} The apparent ambiguity reflected in the non-trivial interval for values of prevalence after a negative test result (cf.\ \autoref{tab:info_pcr}) or NPV (cf.\ \autoref{proposition:npv_y}) is only a manifestation of the ambiguity about prevalence, but it is not due to the test itself.

\subsection{The new test} \label{subsection:newtest}
Next, the new test's informativeness is analyzed. First, I will discuss informativeness only based on the tested population. For this subpopulation 
the prevalence is given by $\overline \chi=\nicefrac{\gamma}{\sigma}$ and therefore the ambiguity about prevalence is muted. \autoref{subsection:overall} extends the analysis then to the informativeness of the new test for the overall population. For the test population, the relevant measures are again positive-predictive value (PPV) and negative-predicative value (NPV), but now they are also conditional on being tested:
\begin{align*}
	\ppv_z &:= P(x=1|z=1, t=1) = \frac{P(x=1,z=1| t=1)}{P(z=1| t=1)} = \frac{P(x=1,z=1| t=1)}{\zeta} \\
	\npv_z &:= P(x=0|z=0, t=1) = \frac{P(x=0,z=0| t=1)}{P(z=0| t=1)} = \frac{P(x=0,z=0| t=1)}{1-\zeta}. 
\end{align*}
To obtain these measures, the distribution $P(x,z| t=1)$ is needed. For a fixed $\tau$, I use a result from \cite{joe-1997} that provides the set of all possible joint distributions $P(x,y,z)$ compatible with the data $P(x,y|t=1)$ (cf.\ \autoref{section:characterization_xyz}). Setting $\tau=1$ in this construction gives the possible distributions $P(x,y,z|t=1)$. Finally, $P(x,z| t=1)$ is obtained by marginalization.

To simplify the algebraic expressions it will be useful to differentiate between four cases defined in \autoref{tab:cases}. Fixing the established test's sensitivity $\sigma$, the test data $P(x,y|t=1)$ immediately reveals the case the test belongs to. \autoref{fig:cases} illustrates this for three real tests considered later (StQ, BiN, CT) and three hypothetical tests (including the dilation test from \autoref{tab:dilation_test}). When $\sigma \rightarrow 1$, then all but the informative case (I) cease to be relevant. For SARS-CoV-2 detecting Antigen test the WHO recommends a minimum specificity close to one. Tests close to the (top-right) frontier in \autoref{fig:cases} satisfy this criterion.\footnote{CT, Uni, and Anti do not satisfy the WHO minimum requirement of a $97\%$ minimum specificity.} Thus, for most applications, either the \emph{confirmatory} (if $\sigma < 1$) or the \emph{informative} case (if $\sigma  \approx 1$) will be the relevant ones.
\begin{table}[ht!]
	\centering
	\caption{Cases relating the two tests}
	\label{tab:cases}
		\begin{tabular}{@{}cc@{}}
		\toprule
		Case name & Parameter restriction	\\ \midrule
		\emph{Confirmatory (C)} 
			& $P(y=0,z=0|t=1) \geq  \max\left\{\overline{\chi}(1-\sigma),1-\overline{\chi}\right\}$\\[1ex]
		\emph{Informative (I)} 
			& $1-\overline{\chi} > P(y=0,z=0|t=1)\geq  \overline{\chi}(1-\sigma)$ \\[1ex]
		\emph{Uninformative (U)} 
			& $\overline{\chi}(1-\sigma) > P(y=0,z=0|t=1) \geq 1-\overline{\chi} $\\[1ex]
		\emph{Contradictory (X)} 
			& $\min\left\{\overline{\chi}(1-\sigma),1-\overline{\chi}\right\} > P(y=0,z=0|t=1)$\\
		\bottomrule
		\multicolumn{2}{c}{\parbox{13cm}{\vspace*{0.25em} \centering \scriptsize Recall $\overline{\chi}=\gamma/\sigma$ is the upper bound on prevalence and $\gamma = P(y=1|t=1)$ is established test's yield.}}
	\end{tabular}
\end{table}

\newcommand\sig{0.75}
\begin{figure}[ht!]\centering
	\begin{tikzpicture}[line cap=round, line join=round,
		x=5cm, y=5.0cm, scale=1.5,
		dot/.style={circle,fill=black,minimum size=4pt,inner sep=0pt,
			outer sep=-1pt}]
		\draw [->,color=black] (-0.025,0) -- (1.05,0) node [right] {$\gamma$};
		\draw [->,color=black] (0,-0.025) -- (0,1.05) node [above] {$P(y=0,z=0|t=1)$};
		\draw [-,color=black] (\sig,0.015) -- (\sig,-0.015) node [below] {\footnotesize $\sigma$};
		\draw [-,color=black] (1,0.015) -- (1,-0.015) node [below] {\footnotesize $1$};
		\draw [-,color=black] (0.015,1-\sig) -- (-0.015,1-\sig) node [left] {\footnotesize $1-\sigma$};
		\draw [-,color=black] (0.015,1) -- (-0.015,1) node [left] {\footnotesize $1$};
		\draw [color=black] (0,0) node [below left] {\footnotesize $0$};
		\draw [color=black, thick] (0,0) -- (\sig,0) -- (\sig, 1-\sig) -- (0,1) -- (0,0);
		\draw[blue,pattern=north east lines,pattern color=blue, opacity=0.4] (0,0) -- (\sig, 1-\sig) -- (0,1) -- (0,0);
		\draw[red,pattern=dots,pattern color=red] (0,1) -- (\sig, 0) -- (\sig, 1-\sig) -- (0,1);
		\path (0.3611,0.6371) node[dot,label=above right:{StQ}]  {};
		\path (0.593,0.395) node[dot,label=above right:{Dil}] {};
		\path (0.2543,0.7348) node[dot,label=above right:{BiN}] {};
		\path (0.4301,0.3886) node[dot,label=above right:{CT}] {};
		\path (0.5,0.25) node[dot,label=left:{Uni}] {};
		\path (0.5,0) node[dot,label=below:{Anti}] {};
		\path (\sig/3,0.3) node[label={(I)}] {};
		\path (\sig/1.675,0.425) node[label={(C)}] {};
		\path (\sig/1.625,0.0) node[label={(X)}] {};
		\path (\sig*1.033,0.15) node[label=left:{(U)}] {};
	\end{tikzpicture}
	\caption[Illustrating the four cases.]{Illustrating the four cases with $\sigma = \sig$:\\{\scriptsize \emph{Uni} is a test corresponding to a uniform distribution $P(y,z|t=1)=1/4$.  \emph{Anti} always produces the opposite result of the established test $P(y=1,z=0|t=1)=P(y=0,z=1|t=1)=1/2$. \emph{StQ}, \emph{BiN}, and \emph{CT} are real tests studied later in \autoref{section:appliaction}. \emph{Dil} is the dilation test given by \autoref{tab:dilation_test}.}}\label{fig:cases}
\end{figure}
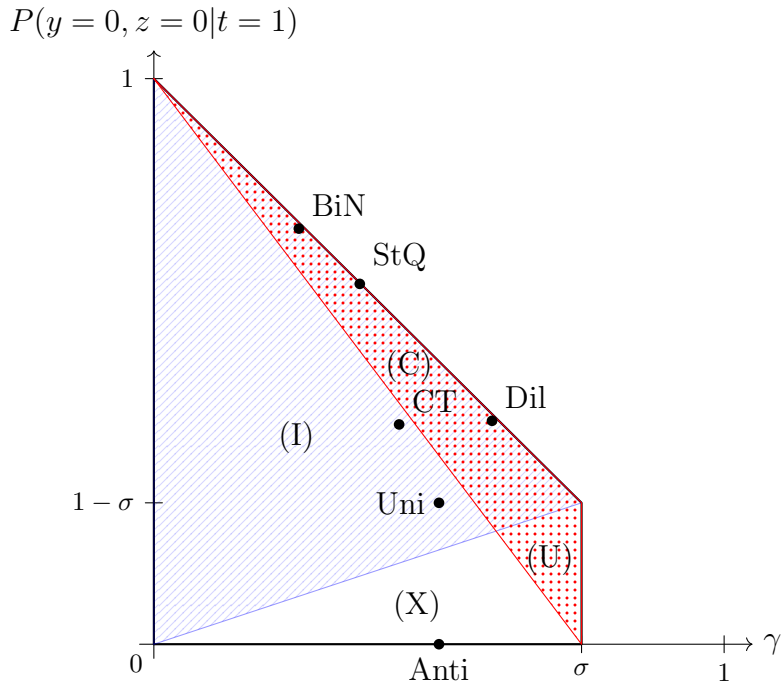

In contrast to the established test, the new test's PPV could be less than one and is, in general, only set-identified. The reason for set-identification is that $P(x,z| t=1)$ is not directly observed. As explained above, there are multiple distributions $P(x,z| t=1)$ consistent with the data and each distribution leads to a potentially different PPV. \autoref{proposition:bounds_ppv} establishes the sharp identified set for values of PPV. 
\begin{proposition}[PPV] \label{proposition:bounds_ppv}
	Under \autoref{assumption:non-trivial}, \autoref{assumption:no_false-pos}, and \autoref{assumption:health-suff}, the new test's positive predictive value is \begin{align*}
		\ppv_z \in 
		\begin{cases}
			\left[P(y=1| z=1,t=1), 1\right]		&\mbox{in case } \text{(C)} \\
			\left[P(y=1| z=1,t=1), P(y=1|z=1,t=1) + \overline{\chi}\frac{1-\sigma}{\zeta}\right]	&\mbox{in case } \text{(I)} \\
			\left[1-\frac{1-\overline{\chi}}{\zeta}, 1\right]	&\mbox{in case } \text{(U)} \\
			\left[1-\frac{1-\overline{\chi}}{\zeta}, P(y=1|z=1,t=1) +\overline{\chi}\frac{1-\sigma}{\zeta}\right]	&\mbox{in case } \text{(X)}.
		\end{cases}
	\end{align*}
\end{proposition}

\begin{proof}
	Conditional on $t=1$ is the same as if $\tau=1$. Thus,
	From \autoref{tab:triple_pmf_lower} and \autoref{tab:triple_pmf_upper}, we obtain respectively:\footnote{Here and in the following I will use $\underline{P}$ to denote lower bound distributions and $\overline P$ for upper bounds.}
	\begin{align*}
		\underline{P}(x=1, z=1|t=1) &:= P(y=1, z=1|t=1)+\max\left\{0, \underbrace{P(y=0, z=1|t=1)-(1-\overline{\chi})}_{=\overline{\chi} - \gamma - P(y=0, z=0|t=1)}\right\} \\
		\overline{P}(x=1, z=1|t=1) &:= P(y=1, z=1|t=1)+\min\left\{\overline{\chi},1-P(y=0, z=0|t=1) \right\} - \gamma \\
		&= \min\left\{\overline{\chi},1-P(y=0, z=0|t=1) \right\} - P(y=1, z=0|t=1) \\
		&= \min\left\{\underbrace{\overline{\chi} - P(y=1, z=0|t=1)}_{=\overline{\chi} - \gamma + P(y=1, z=1|t=1)} ,\zeta \right\}.
	\end{align*} 
	Now, note that $\overline{\chi}- \gamma = \overline{\chi}(1-\sigma)$ and divide by $\zeta$ to obtain PPV.
\end{proof}

To avoid partially identified predictive values, these measures for the new tests are often reported \emph{as if} the reference test is perfect. In this case, the data $P(y,z)$ alone delivers a unique predictive value:
\begin{corollary}[Perfect Gold Standard - PPV] \label{corollary:perfect_GS_PPV}
	Suppose \autoref{assumption:non-trivial}, \autoref{assumption:no_false-pos}, and \autoref{assumption:health-suff} hold. If $\sigma =1$, i.e. the established test has perfect sensitivity, then
	$
		PPV_z = P(y=1|z=1, t=1)
	$.
\end{corollary}

\begin{proof}
	If $\sigma=1$, then $\overline{\chi}= \gamma = P(y=1|t=1)$. Thus, the relevant case is (I).\footnote{Technically, one could be on the border of case (C) or (U), but by continuity the resulting bounds do not change.} Therefore the lower bound is $P(y=1|z=1, t=1)$ and the upper bound is
	$
		P(y=1|z=1, t=1) + \underbrace{\frac{\gamma}{\zeta}\frac{1-\sigma}{\sigma}}_{=0}.
	$
\end{proof}

We saw before that the established test always achieves a maximal PPV of one and therefore provides a lot of information in case it delivers a positive result. How informative is a positive result of the new test? To answer this question, note that since we condition on being tested, there is no prior uncertainty as the prevalence in the testing pool is given by $\overline{\chi}$. Even without this prior ambiguity there remains ambiguity in the test result. For example, in the confirmatory case (C) the interval's width of possible values for $\ppv_z$ is $1-P(y=1|z=1, t=1)$, which is usually small---but non-zero---in applications.	Thus, the information obtained from a new test is ambiguous at least after a positive test result.\footnote{This observation alone implies the test is \emph{not} an experiment \`{a} la \cite{blackwell-1951}. Similarly to deriving PPV, it is possible to derive the new test's sensitivity. This will be a set in general too. Thus, there is a correspondence from (health) states to distributions over signals (test results).}

In contrast to the established test as discussed in \autoref{subsection:establishedtest}, the ambiguity arising from the new test allows for the occurrence of dilation. In the current setting a dilation occurs if $\overline{\chi} = P(x=1|t=1)$ is contained in the intersection
of the two sets with possible values for $P(x=1|y=i, t=1)$ for each test result $i \in \{0,1\}$. Is it possible that after a positive test result the set of possible values for $P(x=1|y=1, t=1)$ contain $\overline{\chi}=P(x=1|t=1)$? \autoref{corollary:ppv_dilation} provides a full characterization. The corresponding case after a negative test result will be discussed afterwards.

\begin{corollary}[Bounds Increase - PPV] \label{corollary:ppv_dilation}
	Suppose \autoref{assumption:non-trivial}, \autoref{assumption:no_false-pos}, and \autoref{assumption:health-suff} hold. The new test's possible values for PPV contain $\overline{\chi}=P(x=1|t=1)$  if and only if
	\begin{align*}
		\sigma \leq \min\left\{\frac{\gamma (1-\zeta)}{P(y=1, z=0|t=1)}, \frac{\gamma \zeta}{P(y=1, z=1|t=1)}\right\},
	\end{align*}
	where the first entry corresponds to an increase in the upper bound, and the second condition ensures the lower bound decreases.
\end{corollary}

\begin{proof}
	For the upper bounds, note that a strict decrease can only happen if and only if cases (I) or (X) occur (i.e. $1-\overline{\chi}> P(y=0, z=0|t=1)$) and  $P(y=1|z=1,t=1) + \overline{\chi}\frac{1-\sigma}{\zeta} < \overline{\chi}$. The first is equivalent to $\sigma > \frac{\gamma}{1-P(y=0, z=0|t=1)}$ and the second to $\sigma>\frac{1-\zeta}{P(z=0|y=1,t=1)}=\frac{\gamma}{P(y=1|z=0,t=1)}$. Since $P(y=0, z=0|t=1) \leq P(y=0| z=0,t=1)$ we also have $P(y=1| z=0,t=1) \leq  1-P(y=0, z=0|t=1)$. Thus, a strict decrease happens if and only if $\sigma > \frac{\gamma}{P(y=1|z=0,t=1)}$.
	
	For the lower bound, note that $1-\frac{1-\overline{\chi}}{\zeta} \leq \overline{\chi}$ always holds. For the other cases (C and I, i.e. $P(y=0, z=0|t=1)\geq\overline{\chi}(1-\sigma)$), a strict increase is $P(y=1| z=1,t=1)> \chi$, which is equivalent to $\sigma > \frac{\gamma}{P(y=1| z=1,t=1)}$. Whereas the case condition is equal to $\sigma \geq \frac{\gamma}{1-P(y=0, z=1|t=1)}$. Similar to above,  $P(y=1| z=1,t=1) \leq  1-P(y=0, z=1|t=1)$ and therefore a strict increase happens if and only if $\sigma > \frac{\gamma}{P(y=1| z=1,t=1)}$.
\end{proof}

%
The inequality of \autoref{corollary:ppv_dilation} becomes non-trivial in case the testing data does not correspond to an independent distribution, which will be the case for most applications. In these cases, a dilation cannot occur when the non-trivial inequality of \autoref{corollary:ppv_dilation} is violated.

It remains to analyze the information contained in a negative test result. \autoref{proposition:bounds_npv} establishes sharp bounds for the negative predictive value of the new test. This is the relevant measure for analyzing how informative a negative test result is.
\begin{proposition}[NPV] \label{proposition:bounds_npv}
		Under \autoref{assumption:non-trivial}, \autoref{assumption:no_false-pos}, and \autoref{assumption:health-suff}, the new test's negative predictive value is \begin{align*}
		\npv_z \in 
		\begin{cases}
			\left[P(y=0|z=0, t=1)		- \overline{\chi}\frac{1-\sigma}{1-\zeta}, \frac{1 - \overline{\chi}}{1-\zeta}\right]		&\mbox{in case } \text{(C)} \\
			\left[P(y=0|z=0, t=1)		- \overline{\chi}\frac{1-\sigma}{1-\zeta}, P(y=0|z=0, t=1)\right]	&\mbox{in case } \text{(I)} \\
			\left[0, \frac{1 - \overline{\chi}}{1-\zeta}\right]	&\mbox{in case } \text{(U)} \\
			\left[0, P(y=0|z=0, t=1)\right]	&\mbox{in case } \text{(X)}.
		\end{cases}
	\end{align*}
\end{proposition}

\begin{proof}
	Conditional on $t=1$ is the same as if $\tau=1$. Thus,
	From \autoref{tab:triple_pmf_lower} and \autoref{tab:triple_pmf_upper}, we obtain respectively:
	\begin{align*}
		\underline{P}(x=0, z=0|t=1) &= \max\left\{0, 1-\overline{\chi} -P(y=0, z=1|t=1) \right\}  \\
		&= \max\left\{0, \underbrace{\gamma-\overline{\chi}}_{=-\overline{\chi}(1-\sigma)} +P(y=0, z=0|t=1) \right\}  \\
		\overline{P}(x=0, z=0|t=1) &= \min\left\{1-\overline{\chi}, P(y=0, z=0|t=1)\right\}.
	\end{align*}
	Division by $P(z=0|t=1)=1-\zeta$ gives NPV.
\end{proof}

The uninformative (U) and contradictory (X) case seem problematic in light of \autoref{proposition:bounds_npv}. In both cases, the lower bound is zero and also the width of the interval is rather large. This is another indication that any reasonable test should not fall in either of these two cases. However, even for the other cases---and like for PPV---the NPV is generally only set-identified. Therefore a negative test result also produces ambiguous information.

Avoiding this ambiguity can be achieved with a perfect reference test. \autoref{corollary:perfect_GS_NPV} verifies that if the reference test is perfect, \autoref{proposition:bounds_npv} reduces to the expression used in many applications and can be calculated directly from the data $P(y,z)$.
\begin{corollary}[Perfect Gold Standard - NPV] \label{corollary:perfect_GS_NPV}
	Suppose \autoref{assumption:non-trivial}, \autoref{assumption:no_false-pos}, and \autoref{assumption:health-suff} hold. If $\sigma =1$, i.e. the established test has perfect sensitivity, then
	$
	\npv_z = P(y=0|z=0, t=1)
	$.
\end{corollary}

\begin{proof}
	If $\sigma=1$, then $\frac{\gamma}{1-\zeta}\frac{1-\sigma}{\sigma}=0$ and as in the proof of \autoref{corollary:perfect_GS_PPV} the relevant case is (I). 
\end{proof}

If there is no perfect reference test available, the negative new test's result leads to ambiguity. Similar to the case of a positive test result, this ambiguity allows for the occurrence of dilation. Using \autoref{proposition:bounds_npv}, \autoref{corollary:npv_dilation} provides a characterization of when the set of possible values of $P(x=1|z=0,t=1)=1-\npv_z$ contains the prior information $P(x=1|t=1) = \overline{\chi}$.
\begin{corollary}[Bounds Increase - NPV] \label{corollary:npv_dilation}
	Suppose \autoref{assumption:non-trivial}, \autoref{assumption:no_false-pos}, and \autoref{assumption:health-suff} hold. The set of possible values for $P(x=1|z=0,t=1)$ includes the prior information $\overline{\chi}=P(x=1|t=1)$ if and only if 
	\begin{align*}
	\sigma \leq \min\left\{ \frac{\gamma \zeta}{P(y=1, z=1|t=1)}, \frac{\gamma (1-\zeta)}{P(y=1, z=0|t=1)}\right\},
	\end{align*}
where the first entry corresponds to an increase in the upper bound, and the second condition ensures the lower bound decreases.
\end{corollary}

\begin{proof}
	For the upper bound (of $1-\npv_z$) a strict decrease can only happen in cases (C) and (I), i.e. $P(y=0, z=0|t=1)\geq\overline{\chi}(1-\sigma)$.
	In these cases, a strict decrease is equivalent to $(1-\zeta) -  \left[P(y=0, z=0|t=1)-\chi(1-\sigma)\right] < \chi (1-\zeta)$ or $1-\zeta + \overline{\chi}(\zeta-\sigma) < P(y=0, z=0|t=1) = 1 - \zeta - P(y=1, z=0|t=1)$. Rearranging gives,
	\begin{align*}
		\sigma > \frac{\zeta}{P(z=1|y=1,t=1)} = \frac{\gamma\zeta}{P(y=1,z=1|t=1)}.
	\end{align*}
	As in the proof of \autoref{corollary:ppv_dilation} the condition for being in case (C) or (I) 	is implied by this condition.
	
	For the lower bound, a decrease occurs if 
	\begin{align*}
		\chi& (1-\zeta) \geq \\
		&(1-\zeta) - \min\left\{1-\chi, P(y=0, z=0|t=1)\right\} = \max\left\{\chi-\zeta, P(y=1, z=0|t=1)\right\}.
	\end{align*}
	 First,  $\chi-\zeta \leq \chi (1-\zeta)$ always holds as  $\chi \leq 1$. Second, rearranging $P(y=1, z=0|t=1) \leq (1-\zeta)\frac{\gamma}{\sigma}$ provides the condition.
\end{proof}

\autoref{corollary:ppv_dilation} combined with \autoref{corollary:npv_dilation} provides an exact characterization for when the new test is a dilation. In fact, as the conditions are the same a dilation occurs if and only if 
\begin{align}
	\sigma \leq \min\left\{ \frac{\gamma \zeta}{P(y=1, z=1|t=1)}, \frac{\gamma (1-\zeta)}{P(y=1, z=0|t=1)}\right\}. \label{eq:dilation}
\end{align}
When evaluating a new test's accuracy it is important to make sure the data violates \autoref{eq:dilation}. Otherwise, the test is uninformative in an extreme sense. In typical applications, the data often satisfies $P(y=1, z=0|t=1)\leq\gamma (1-\zeta)$.\footnote{Even data that regards a test as inadequate as in \cite{cassaniti-etal-2020} satisfies this inequality. I thank Filip Obradovic for making me aware of this report.}  test   In these cases, a dilation can only occur if $\sigma \leq \frac{\gamma \zeta}{P(y=1, z=1|t=1)}$.

The \cite{who-2020} recommends minimum quality requirements using only information directly provided by the data $P(y,z|t=1)$. In light of this analysis, an evaluation should also take $\sigma$, the established test's sensitivity, into account and with this also make sure that the test is not a dilation. $\sigma \leq \frac{\gamma \zeta}{P(y=1, z=1|t=1)}$ combined with a given minimum standard provides an easy-to-verify sufficient condition to avoid dilation.

For this, let $\underline{\Sigma}$ be a minimum (apparent) sensitivity threshold below which a test is deemed not reliable and denote the new test's apparent sensitivity with $\Sigma = P(z=1|y=1, t=1)$, so that a test is reliable if $\Sigma > \underline{\Sigma}$.\footnote{Usually, the minimum requirements include also a threshold for specificity, but this does not matter here.} Then, the application relevant case from \autoref{eq:dilation} to avoid a dilation can be expressed as $\sigma > \zeta /\Sigma$ or equivalently as $\Sigma > \zeta/\sigma$. If $\underline{\Sigma} \geq \zeta/\sigma$, then any test meeting the minimum requirement cannot be a dilation. Thus, it suffices to make sure the new test's yield is not too high:\footnote{It is worth recalling that this is a sufficient condition when, additionally, $\frac{\gamma \zeta}{P(y=1, z=1|t=1)} \leq \frac{\gamma (1-\zeta)}{P(y=1, z=0|t=1)}$ and in many applications this inequality becomes irrelevant for \autoref{eq:dilation} because the right-hand side is greater than one.}
\begin{align}
	\zeta := P(z=1|t=1) \leq \sigma \times \underline{\Sigma}. \label{eq:sufficient_no_dilation}
\end{align}
If the established test is highly specific, i.e. $\sigma \approx 1$, then \autoref{eq:sufficient_no_dilation} is satisfied unless the new test's yield is extremely high.

For SARS-CoV-2 Antigen tests the WHO recommendation is $\underline{\Sigma} = 0.8$ and if the PCR test is not highly specific then \autoref{eq:sufficient_no_dilation} might be violated. For example, the dilation test of \autoref{tab:dilation_test} has $\zeta=0.49$ and if the PCR has sensitivity of $\sigma=0.6$ then not only \autoref{eq:sufficient_no_dilation} is violated but the test is a dilation. More specifically, \autoref{eq:dilation} can be used to find the exact threshold sensitivity $\sigma^*$ below which a given test turns into a dilation. For the dilation test this value is $\sigma^* = 60.79\%$.

\subsection{Informativeness of Additional Testing} \label{subsection:additest}
Whereas the new test produces ambiguous information, the established test is always informative. Therefore, practitioners might want to perform an additional established test depending on whether a person obtains a negative or positive result from the new test. For example, if an Antigen test is used to detect SARS-CoV-2 and the result is positive, a common practice is verifying the result by means of a PCR test. Since PCR tests are the reference test for evaluating the accuracy of Antigen tests, the current framework can be used to shed light on how informative this additional test is.

\begin{proposition}[Combined testing] \label{proposition:both_tests}
	Under \autoref{assumption:non-trivial}, \autoref{assumption:no_false-pos}, and \autoref{assumption:health-suff},
	\begin{align*}
		P(x=1|y=1,z=1,t=1)=P(x=1|y=1,z=0,t=1)=1,
	\end{align*}
	\begin{align*}
		P(x=0|y=0,z=0,t=1) \in
		\begin{cases}
			\left[1-\frac{\overline{\chi}(1-\sigma)}{P(y=0, z=0|t=1)}, 1\right]		&\mbox{in case } \text{(C)} \\
			\left[1-\frac{\overline{\chi}(1-\sigma)}{P(y=0, z=0|t=1)}, \frac{1-\overline{\chi}}{P(y=0, z=0|t=1)}\right]	&\mbox{in case } \text{(I)} \\
			\left[0,1\right]	&\mbox{in case } \text{(U)} \\
			\left[0, \frac{1-\overline{\chi}}{P(y=0, z=0|t=1)}\right]	&\mbox{in case } \text{(X)},
		\end{cases}
	\end{align*}
and
	\begin{align*}
		P(x=0|y=0,z=1,t=1) \in
		\begin{cases}
			\left[0, 1\right]		&\mbox{in case } \text{(C)} \\
			\left[0, \frac{1-\overline{\chi}}{P(y=0, z=1|t=1)}\right]	&\mbox{in case } \text{(I)} \\
			\left[1-\frac{\overline{\chi}(1-\sigma)}{P(y=0, z=1|t=1)},1\right]	&\mbox{in case } \text{(U)} \\
			\left[1-\frac{\overline{\chi}(1-\sigma)}{P(y=0, z=1|t=1)}, \frac{1-\overline{\chi}}{P(y=0, z=1|t=1)}\right]	&\mbox{in case } \text{(X)}.
		\end{cases}
	\end{align*}
\end{proposition}

\begin{proof}
	If $y=1$ the the PPV has to be one independent of the new test's result because of \autoref{assumption:no_false-pos}. For NPV, again start from \autoref{tab:triple_pmf_lower} and \autoref{tab:triple_pmf_upper} with $\tau=1$. First, the case of both tests matching, i.e. $y=0=z$:
	\begin{align*} 
		\underline{P}(x=0, y=0, z=0|t=1) &= \max\left\{0, \gamma-\overline{\chi}+P(y=0, z=0|t=1)\right\} \\
		&= \max\left\{0, P(y=0, z=0|t=1) -\overline{\chi}(1-\sigma)\right\} \\
		&\text{and } \\
		\overline{P}(x=0, y=0, z=0|t=1)&=\min\left\{1-\overline{\chi}, P(y=0, z=0|t=1)\right\}.
	\end{align*}
	Now, divide by $P(y=0, z=0|t=1)$ to obtain $P(x=0|y=0,z=0,t=1)$.
	
	In case of differing test results, the relevant probability are:
	\begin{align*}
		\underline{P}(x=0, y=0, z=1|t=1) &= \min\left\{1-\frac{\gamma}{\sigma}, P(y=0, z=1|t=1)\right\} \\
		&=\min\left\{1-\frac{\gamma}{\sigma}, 1-\gamma - P(y=0, z=0|t=1)\right\} \\
		&= 1-\gamma - \max\left\{\overline{\chi}(1-\sigma), P(y=0, z=0|t=1)\right\} \\
		&\text{and } \\
		\overline{P}(x=0, y=0, z=1|t=1)&=\max\left\{1-\overline{\chi}-P(y=0, z=0|t=1),0 \right\} \\
		&=\max\left\{P(y=0, z=1|t=1) - \overline{\chi}(1-\sigma),0 \right\}.
	\end{align*}
	Note that $\underline{P}(x=0, y=0, z=1|t=1) \geq \overline{P}(x=0, y=0, z=1|t=1)$ in this case. Division by $P(y=0, z=1|t=1)$ gives $P(x=0|y=0,z=1,t=1)$.
\end{proof}

\autoref{proposition:both_tests} once more reveals that tests in the category (U) and (X) should be avoided. Even if the two test results match and are both negative, the possibility of zero (negative) predictive value cannot be ruled out. \autoref{proposition:both_tests} also makes clear the naming convention of the cases defined in \autoref{tab:cases}. A confirmatory test (C) provides accurate information when both test produce a negative result, but is completely uninformative if and only if the new test has a positive result. An informative test (I), however, always provides some information in the sense of producing not completely trivial bounds. A contradictory test (X) provides information, but leans against the result of the established test. The uninformative test (U) provides no information at all even when both tests agree on a negative result. Of course, performing the additional test is always informative in the sense of not being a dilation. The established test does not produce false-negatives and therefore a positive result from the established test is always a perfect predictor of being infected regardless of the new test.

\subsection{Predictive values for the overall population} \label{subsection:overall}
\autoref{proposition:npv_y} bounds the established test's NPV for the overall population, not only for the tested population. The new test, on the other hand, was analyzed for the testing pool only so far. The full characterization in \autoref{section:characterization_xyz} allows to extend the analysis of the new test to make an evaluation for the overall population. Since this involves more cumbersome notation, I only illustrate the resulting bounds for the $\npv = P(x=0|z=0)$. The analysis of PPV would proceed in a similar matter.

\begin{proposition}[Unconditional NPV] \label{proposition:bounds_npv_overall}
	Under \autoref{assumption:non-trivial}--\autoref{assumption:health-suff}, the new test's (unconditional) negative predictive value is sharply bounded by \begin{align*}
			&\left[\frac{1-\overline{\chi} -\tau P(y=0, z=1|t=1)}{1-\tau\zeta}, \frac{1 - \underline{\chi}}{1-\tau\zeta}\right]		&\mbox{in case } &\text{(C*)} \\
			&\left[\frac{1-\overline{\chi} -\tau P(y=0, z=1|t=1)}{1-\tau\zeta}, 1-\frac{P(y=1, z=0|t=1)}{1-\tau\zeta}\right]	&\mbox{in case } &\text{(I*)} \\
			&\left[0, \frac{1 - \underline{\chi}}{1-\tau\zeta}\right]	&\mbox{in case } &\text{(U*)} \\
			&\left[0,  1-\frac{P(y=1, z=0|t=1)}{1-\tau\zeta}\right]	&\mbox{in case } &\text{(X*)},
	\end{align*}
where
\begin{align*}
	\text{(C*)} \quad &\ldots \quad P(y=0, z=0|t=1) \geq \max\left\{1-\gamma -\frac{1-\overline{\chi}}{\tau},1-\overline{\chi}\right\} \\
	\text{(I*)} \quad &\ldots \quad 1-\overline{\chi} > P(y=0, z=0|t=1) \geq 1-\gamma -\frac{1-\overline{\chi}}{\tau} \\
	\text{(U*)} \quad &\ldots \quad 1-\gamma -\frac{1-\overline{\chi}}{\tau} > P(y=0, z=0|t=1) \geq  1-\overline{\chi} \\
	\text{(X*)} \quad &\ldots \quad \min\left\{1-\gamma -\frac{1-\overline{\chi}}{\tau},1-\overline{\chi}\right\} > P(y=0, z=0|t=1).
\end{align*}
\end{proposition}

\begin{proof}
	From \autoref{tab:triple_pmf_lower} and \autoref{tab:triple_pmf_upper}, we obtain respectively:
	\begin{align*}
		\underline{P}(x=0, z=0) &= \max\left\{0, 1-\overline{\chi} -\tau P(y=0, z=1|t=1) \right\}  \\
		\overline{P}(x=0, z=0) &= 1-\tau\max\left\{\overline{\chi}, 1-P(y=0, z=0|t=1)\right\} \\
		&= 1-\tau\max\left\{\overline{\chi}, \zeta+P(y=1, z=0|t=1)\right\} .
	\end{align*}
Now, the result follows from dividing by $P(z=0) = 1 -\tau\zeta$.
\end{proof}

If instead of predictive values the interest lies in the new test's sensitivity or specificity in the whole population another complication arises. Conditional on the testing pool, both of these measures can be derived as in \autoref{subsection:newtest}. For example, for sensitivity one could use the proof of \autoref{proposition:bounds_ppv} and divide by $P(x=1|t=1)=\overline{\chi}$ instead of $P(z=1|t=1)=\zeta$. The bounds for sensitivity are again determined by considering the extremes of \autoref{tab:triple_pmf_lower} and \autoref{tab:triple_pmf_upper}. For the unconditional sensitivity, however, the the numerator and the denominator are both set identified because $P(x=1) \in [\underline{\chi}, \overline{\chi}]$. Therefore, the lower bound might not be attained at either of the extreme distributions. This makes solving for a closed-form expression for sensitivity intractable. Nonetheless, the bounds can easily be obtained computationally by considering a fixed $\Gamma := P(y=1) \in [\tau \gamma, \gamma]$ with corresponding $p =\Gamma/\sigma$. For this $\Gamma$, sharp bounds of sensitivity, say $[L_\Gamma, H_\Gamma]$, can be obtained by using \autoref{tab:triple_cdf_lower_Gamma} and \autoref{tab:triple_cdf_upper_Gamma}. To find the overall bounds for sensitivity, two (non-linear) optimization problems across all values of $\Gamma$ need to be performed to give $[\min_\Gamma L_\Gamma, \max_\Gamma  H_\Gamma]$.

\section{Applications} \label{section:appliaction}
In this section, the theoretic framework will be illustrated with several applications. First, I analyze the (hypothetical) dilation test presented in the introduction. Then, I examine two real SARS-CoV-2 detecting tests. Finally, I show that CT-scanning procedures to detect COVID-19 are prone to being dilations.
\subsection{A Dilation Test}
As argued before the hypothetical test data in \autoref{tab:dilation_test} corresponds to a dilation. Suppose the test data is derived for an Antigen test to detect SARS-CoV-2 and the reference test is a PCR test.\footnote{Recall that for SARS-CoV-2 detection a PCR test is the established test used to evaluate other tests. \citep{esbin-etal-2020}\label{footnote:PCR_GS}}. The test satisfies the \citeauthor{who-2020}'s (\citeyear{who-2020}) minimum requirements with apparent sensitivity ($\Sigma=80.6\%$) and specificity ($97.1\%$) above the specified thresholds of $80\%$ and $97\%$, respectively.\footnote{These numbers are calculated as if the reference test is perfect. This is similar to \autoref{corollary:perfect_GS_PPV} and \autoref{corollary:perfect_GS_NPV}.} For such a setting the current framework is applicable. Especially, \autoref{assumption:no_false-pos} seems to be warranted because a PCR test is highly specific. However, it is known that a PCR test might lack high sensitivity. \citet{alcoba-florez-etal-2020} report sensitivity for several PCR tests with point estimates ranging from $\sigma=60.2\%$ to $\sigma=97.9\%$.\footnote{\citeauthor{alcoba-florez-etal-2020} differentiate values based on the targeted gene.  The range reported here is across all genes and tests.} All of the $95\%$ confidence intervals exclude perfect sensitivity,  $\sigma=1$.

Using the results from \autoref{subsection:newtest}, \autoref{tab:dilation} summarizes some key statistics for the dilation test. When the PCR sensitivity is close to one, the new (hypothetical) test produces relative accurate measurements with PPV close to one and NPV above $75\%$. However, if the PCR test lacks high sensitivity then we cannot be sure of the dilation test's quality. In the worst-case for PCR sensitivity ($\sigma=0.6$), the new test is indeed a dilation: Before a test result was obtained the prevalence (in the testing pool) is $98.8\%$, after obtaining \emph{either} dilation test's result the possible probability of being infected is at least the interval $[97.7\%, 100\%]$. In fact, potentially even more puzzling is that the lowest value after a negative test is strictly higher than after a positive result. Using \autoref{eq:dilation}, $\sigma^*=60.8\%$ represents the cutoff PCR sensitivity below which a dilation occurs.
\begin{table}[ht!]
	\centering
	\caption{Dilation Test Statistics}
	\label{tab:dilation}
	\begin{tabular}{@{}ccccc@{}}
		\toprule
		$\sigma$ & $0.6$ & $0.85$ & $0.98$	& $1$\\ \midrule
		$\overline{\chi}=P(x=1|t=1)$ & $98.8\%$ & $69.8\%$ & $60.5\%$ & $59.3\%$ \\[0.5ex]
		$\ppv_z$ & 	$[97.6\%, 100\%]$ &	$[97.6\%, 100\%]$ & $[97.6\%, 100\%]$ & $97.6\%$\\[0.5ex]
		$1-\npv_z$ & $[97.7\%, 100\%]$ &	$[40.7\%, 43.1\%]$ & $[22.6\%, 24.9\%]$ & $22.6\%$\\[0.5ex]
		$\npv_z$ & $[0\%, 2.29\%]$ &	$[56.9\%, 59.3\%]$ & $[75.1\%, 77.43\%]$ & $77.43\%$\\[0.5ex]
		Dilation Threshold & \multicolumn{4}{c}{$\sigma^*=60.8\%$} \\
		\bottomrule
	\end{tabular}
\end{table}

\subsection{Standard Q COVID-19 Rapid Antigen Test}
Next, consider the Standard Q (\emph{StQ}) COVID-19 Rapid Antigen Test of SD Biosensor/Roche for detection of SARS-CoV-2 as analyzed by \cite{kaiser-etal-2020}. They use results of PCR tests as comparison (see \autoref{footnote:PCR_GS}). The testing data is summarized in \autoref{tab:standardQ_test}.
		\begin{table}[ht!]
	\centering
	\caption{StQ Test results from \citet[p. 3]{kaiser-etal-2020}}
	\label{tab:standardQ_test}
	\begin{tabular}{@{}cccc@{}}
		\toprule
		$z \backslash y$ & $y=0$ & $y=1$ & Sum	\\ \midrule
		$z=0$ & $63.71\%$ & $3.97\%$ & $67.67\%$\\
		$z=1$ & $0.19\%$ & $32.14\%$ & $32.33\%$\\ \midrule
		Sum & $63.89\%$ & $36.11\%$ \\
		\bottomrule
	\end{tabular}
\end{table}
When $\sigma=1$, then StQ's PPV and NPV  are obtained with \autoref{corollary:perfect_GS_PPV} and \autoref{corollary:perfect_GS_NPV} which yields $99.42\%$ and $94.13\%$, respectively. These are the reported values of \cite{kaiser-etal-2020}. However, as explained above PCR are not perfectly sensitive.\footnote{ \cite{kaiser-etal-2020} use PCR tests targeting $E$ genes, which tend to have higher sensitivity in the analysis of \cite{alcoba-florez-etal-2020}. The lowest reported sensitivity for a PCR test targeting $E$ genes is $65.33\%$.} Thus, to evaluate the StQ test the current framework is applicable.

Focusing first on the testing pool only, \autoref{tab:StQ-acc} summarizes PPV and NPV for different values of PCR sensitivity ($\sigma$) using \autoref{proposition:bounds_ppv} and \autoref{proposition:bounds_npv}. Even if the PCR test lacks high sensitivity, StQ has a close to perfect positive predicative value ($\ppv_z \approx 1$). However, the values for NPV drop significantly as $\sigma$ decreases. In the worst case, a negative StQ result becomes close to a fair coin flip. However, the test is very informative overall as can be seen by the low dilation threshold $\sigma^*=36.3\%$.
\begin{table}[ht!]
	\centering
	\caption{Accuracy of StQ}
	\label{tab:StQ-acc}
	\begin{tabular}{@{}cccc@{}}
		\toprule
		$\sigma$ & $0.6$ & $0.85$ & $0.98$	\\ \midrule
		$\ppv_z$ &  	$[99.4\%, 100\%]$ & $[99.4\%, 100\%]$ & $[99.4\%, 100\%]$\\[0.5ex]
		$\npv_z$ &  	$[58.6\%, 58.9\%]$ & $[84.7\%, 85\%]$ & $[93.1\%, 94.1\%]$\\[0.5ex]
		Dilation Threshold & \multicolumn{3}{c}{$\sigma^*=36.3\%$} \\
		\bottomrule
	\end{tabular}
\end{table}

\citet[p. 1]{kaiser-etal-2020} state  ``study participants were representative of the usual population seeking testing in our center (main testing center in Geneva). The majority were presenting with symptoms compatible with a SARS-CoV2 infection and a minority were asymptomatic but with a known positive contact or were asymptomatic healthcare workers.'' The current framework allows to use the obtained testing data to evaluate StQ's quality for the overall population (of Geneva) as analyzed in \autoref{subsection:overall}. Furthermore, this explanation supports \autoref{assumption:test_monotone}.

\autoref{tab:prevalence} shows bounds on prevalence using the baseline analysis in \autoref{section:main}. For low values of $\tau$, i.e. the testing pool was highly non-representative of the overall population, the width of the intervals is rather wide. However, even the lowest number is close to $2\%$ indicating a thorough spread of the virus in Geneva at the time of testing.\footnote{Note that this is a one time analysis. It does not answer the question of how many people were cumulatively infected by SARS-CoV-2 up to the time of testing.} When testing becomes representative ($\tau \rightarrow 1$) the prevalence converges to the prevalence in the testing pool.
\begin{table}[ht!]
	\centering
	\caption{Bounds on prevalence $p \in [\underline{\chi}, \overline{\chi}]$}
	\label{tab:prevalence}
	\begin{tabular}{@{}cccc@{}}
		\toprule
		$\tau\diagdown \sigma$ & $0.6$ & $0.85$ & $0.98$	\\ \midrule
		$\nicefrac{1}{20}$ &  	$[3.01\%, 60.2\%]$ & $[2.12\%, 42.5\%]$ & $[1.84\%, 36.8\%]$ \\[0.75ex]
		$\nicefrac{1}{10}$ &  	$[6.02\%, 60.2\%]$ & $[4.25\%, 42.5\%]$ & $[3.68\%, 36.8\%]$ \\[0.75ex]
		$\nicefrac{1}{2}$ &  	$[30.1\%, 60.2\%]$ & $[21.2\%, 42.5\%]$ & $[18.4\%, 36.8\%]$\\[0.75ex]
		$\nicefrac{19}{20}$ &  $[57.2\%, 60.2\%]$ & $[40.4\%, 42.5\%]$ & $[35.7\%, 36.8\%]$\\[0.75ex]
		$1$ & $60.2\%$ & $42.5\%$& $36.8\%$  \\
		\bottomrule
	\end{tabular}
\end{table}

At this time, if a Genevese obtains a negative PCR result, what is the probability of her being infected? If testing is not competently representative, a unique number cannot be given. However, \autoref{proposition:npv_y} provides sharp bounds for this case and the results are shown in \autoref{tab:pcr_npv}.
\begin{table}[ht!]
	\centering
	\caption{Bounds on $\npv_y$}
	\label{tab:pcr_npv}
	\begin{tabular}{@{}cccc@{}}
		\toprule
		$\tau\diagdown \sigma$ & $0.6$ & $0.85$ & $0.98$	\\ \midrule
		$\nicefrac{1}{20}$ &  	$[62.3\%, 98.8\%]$ & $[90.0\%, 99.7\%]$ & $[98.9\%, 100\%]$ \\[0.75ex]
		$\nicefrac{1}{10}$ &  	$[62.3\%, 97.5\%]$ & $[90.0\%, 99.3\%]$ & $[98.9\%, 99.9\%]$ \\[0.75ex]
		$\nicefrac{1}{2}$ &  	$[62.3\%, 85.3\%]$ & $[90.0\%, 96.1\%]$ & $[98.9\%, 99.6\%]$\\[0.75ex]
		$\nicefrac{19}{20}$ &  $[62.3\%, 65.2\%]$ & $[90.0\%, 90.8\%]$ & $[98.9\%, 98.9\%]$\\[0.75ex]
		$1$ & $62.3\%$ & $90.0\%$& $98.9\%$  \\
		\bottomrule
	\end{tabular}
\end{table}
How do these PCR results compare to results from StQ? \autoref{tab:npv_standardQ} provides the numbers using \autoref{proposition:bounds_npv_overall}. The lower bounds are significantly lower than for the PCR test. This makes the width of the interval also significantly wider. The widening is a reflection of the combination of the two missing data problems inherit in the testing procedure without a perfect reference test: ($i$) unknown overall prevalence (which also affects PCR's NPV) and ($ii$) missing correlation data (which does not affect the PCR's NPV).

\begin{table}[ht!]
	\centering
	\caption{Bounds on STQ's $\npv_z$ for overall population}
	\label{tab:npv_standardQ}
	\begin{tabular}{@{}cccc@{}}
		\toprule
		$\tau\diagdown \sigma$ & $0.6$ & $0.85$ & $0.98$	\\ \midrule
		$\nicefrac{1}{20}$ &  	$[40.5\%, 96.0\%]$ & $[58.5\%, 96.0\%]$ & $[64.2\%, 96.0\%]$ \\[0.75ex]
		$\nicefrac{1}{10}$ &  	$[41.1\%, 95.9\%]$ & $[59.4\%, 95.9\%]$ & $[65.3\%, 95.9\%]$ \\[0.75ex]
		$\nicefrac{1}{2}$ &  	$[47.4\%, 83.4\%]$ & $[68.5\%, 94.0\%]$ & $[75.2\%, 95.2\%]$\\[0.75ex]
		$\nicefrac{19}{20}$ &  $[57.2\%, 61.8\%]$ & $[82.8\%, 86.1\%]$ & $[90.9\%, 93.8\%]$\\
		\bottomrule
	\end{tabular}
\end{table}

\subsection{BiaxNOW Covid-19 Ag Home Test} \label{subsection: BiN}
The BiaxNOW (\emph{BiN}) Covid-19 Ag Home Test of Abbott is one of the first rapid Antigen tests for use at home which is able to detect the SARS-CoV-2 virus that was emergency approved by the \cite{fda-2020a}. BiN's clinical performance for approval by the FDA was conducted with a PCR test as a reference. The data are shown in \autoref{tab:BiN_test} and \autoref{tab:BiN-acc} shows the implied accuracy measures.

\begin{table}[ht!]
	\centering
	\caption{BiN Test results from \citet[p. 20]{fda-2020a}}
	\label{tab:BiN_test}
	\begin{tabular}{@{}cccc@{}}
		\toprule
		$z \backslash y$ & $y=0$ & $y=1$ & Sum	\\ \midrule
		$z=0$ & $73.48\%$ & $3.91\%$ & $77.39\%$\\
		$z=1$ & $1.09\%$ & $21.52\%$ & $22.61\%$\\ \midrule
		Sum & $74.57\%$ & $25.43\%$ \\
		\bottomrule
	\end{tabular}
\end{table}

\begin{table}[ht!]
	\centering
	\caption{Accuracy of BiN}
	\label{tab:BiN-acc}
	\begin{tabular}{@{}cccc@{}}
		\toprule
		$\sigma$ & $0.6$ & $0.85$ & $0.98$	\\ \midrule
		$\ppv_z$ &  	$[95.2\%, 100\%]$ & $[95.2\%, 100\%]$ & $[95.2\%, 97.5\%]$\\[0.5ex]
		$\npv_z$ &  	$[73.0\%, 74.4\%]$ & $[89.1\%, 90.6\%]$ & $[94.3\%, 94.9\%]$\\[0.5ex]
		Dilation Threshold & \multicolumn{3}{c}{$\sigma^*=26.7\%$} \\
		\bottomrule
	\end{tabular}
\end{table}
Relative to StQ, BiN has significantly lower PPV and also rules out perfect PPV for lower values of PCR specificity. On the other hand, NPV is uniformly greater for BiN compared to StQ. Even for the worst-case PCR sensitivity, BiN's possible NPV values are reasonably high. Furthermore, the dilation threshold is extremely low at  $\sigma^*=26.7\%$.

The \cite{fda-2020a} also provides additional data about BiN results by including the cycle threshold obtained by the PCR test.\footnote{\cite{publichealthengland-2020} explains: ``Cycle threshold (Ct) is a semi-quantitative value that can broadly categorise the concentration of viral genetic material in a patient sample following testing by RT PCR as low, medium or high –- that is, it tells us approximately how much viral genetic material is in the sample. A low Ct indicates a high concentration of viral genetic material, which is typically associated with high risk of infectivity. A high Ct indicates a low concentration of viral genetic material which is typically associated with a lower risk of infectivity.''} \autoref{tab:BiN_test_CT} shows this data.
\begin{table}[ht!]
	\centering
	\caption{BiN Test results from \citet[p. 21]{fda-2020a} with Cycle Threshold Count}
	\label{tab:BiN_test_CT}
	\begin{tabular}{@{}cccc@{}}
		\toprule
		$z \backslash y$ & $Ct < 33$ & $Ct \geq 33$ & Sum	\\ \midrule
		$z=0$ & $16.2\%$ & $6.94\%$ & $23.12\%$\\
		$z=1$ & $9.83\%$ & $67.05\%$ & $76.89\%$\\ \midrule
		Sum & $26.01\%$ & $73.99\%$ \\
		\bottomrule
	\end{tabular}
\end{table}
This is additional data a PCR test produces, which can be used to refine bounds on predictive values of the Antigen test. However, an extension of the current setting is needed, because such additional information is not accounted for in the current setting with binary tests. \autoref{subsection:addtionaldata} discusses a possible extension of the current setting to allow for this additional data.

\subsection{CT scan to detect COVID-19}
\cite{ai-etal-2020} and \cite{gietema-etal-2020} propose using chest CT scans for early identifying COVID-19 in patients. In \citeauthor{gietema-etal-2020}'s study, all COVID-19 symptomatic patients of a single Dutch emergency department have a chest CT scan and a PCR test for detecting SARS-CoV-2. Their study design exactly fits the framework of the current paper: ($i$) non-representative sampling of the testing pool and ($ii$) missing correlation data due to use of an imperfect reference test (with perfect specificity). The testing data are reproduced in \autoref{tab:CT_test}.
\begin{table}[ht!]
	\centering
	\caption{CT scan data from \citet[Table 2]{gietema-etal-2020}}
	\label{tab:CT_test}
	\begin{tabular}{@{}cccc@{}}
		\toprule
		$z \backslash y$ & $y=0$ & $y=1$ & Sum	\\ \midrule
		$z=0$ & $38.86\%$ & $4.66\%$ & $43.52\%$\\
		$z=1$ & $18.13\%$ & $38.34\%$ & $56.48\%$\\ \midrule
		Sum & $56.99\%$ & $43.01\%$ \\
		\bottomrule
	\end{tabular}
\end{table}
Compared to the previously studies Antigen tests, the data for CT scans seems less aligned with the PCR test results. This is an indication that such a CT test is less informative: the dilation threshold of $\sigma^*=63.35\%$ is higher than for the Antigen tests. Thus, this testing procedure is completely uninformative if $\sigma=60\%$---the lowest sensitivity of a PCR test reported by \cite{alcoba-florez-etal-2020}. In this case,  \autoref{tab:CT_npvs} shows (sharp bounds on) population prevalence, PCR NPVs, and CT scan NPVs.
\begin{table}[ht!]
	\centering
	\caption{Implications for overall population from \cite{gietema-etal-2020}.}
	\label{tab:CT_npvs}
	\begin{tabular}{@{}cccc@{}}
		\toprule
		$\tau$\textdownarrow & $p:=P(x=1)$ & $\npv_y$ & $\npv_z$	\\ \midrule
		$\nicefrac{1}{20}$ &  	$[3.58\%, 71.7\%]$ & $[49.7\%, 98.5\%]$ & $[23.4\%, 65.1\%]$ \\[0.75ex]
		$\nicefrac{1}{10}$ &  	$[7.17\%, 71.7\%]$ & $[49.7\%, 97.0\%]$ & $[23.4\%, 65.1\%]$ \\[0.75ex]
		$\nicefrac{1}{2}$ &  	$[35.8\%, 71.7\%]$ & $[49.7\%, 81.7\%]$ & $[23.4\%, 65.1\%]$\\[0.75ex]
		$\nicefrac{19}{20}$ &  $[68.1\%, 71.7\%]$ & $[49.7\%, 54.0\%]$ & $[23.4\%, 65.1\%]$\\
		\bottomrule
	\end{tabular}
\end{table}
The  PCR's (assumed) low sensitivity means that its NPV might be quite low, but it as at least close to $50\%$, irrespective of $\tau$. On the other hand, the CT scan has both a sizable interval and a low lower bound of possible NPVs. Since $\sigma=0.6$ is below the dilation threshold, the CT scan is completely uninformative for the tested population (equivalently for $\tau=1$). Furthermore, this remains true for the overall population if $\tau \geq 1/2$ as shown in \autoref{tab:CT_npvs}. For example, for a non-COVID-indicative CT scan the set of possible infection probabilities $P(x=1|z=0)$ increases relative to the prior information $p$.\footnote{Recall that $P(x=1|z=0)=1-\npv_z$.}

Even more striking is the data of \cite{ai-etal-2020} shown in \autoref{tab:CT_test_Ai}.\footnote{I thank Filip Obradovic for providing this reference.} \citeauthor{ai-etal-2020} also use Chest CT scans to test for COVID-19. Their data is obtained in Wuhan, China and like the study of \cite{gietema-etal-2020} a PCR test is used as a reference. The data reveals a very low apparent specificity but a high apparent sensitivity of $\Sigma = 96.51\%$. \autoref{eq:sufficient_no_dilation} indicates that a high yield of the new test, $\zeta$, might be problematic. Here, this yield is very high with $\zeta=87.57\%$. The problem becomes even more apparent by looking at the dilation threshold, which is high with a value of $\sigma^* = 90.74$. This implies that even if the PCR test is quite sensitive, the CT scan is completely uninformative for the tested people in Wuhan.
\begin{table}[ht!]
	\centering
	\caption{CT scan data from \citet[Table 2]{ai-etal-2020}}
	\label{tab:CT_test_Ai}
	\begin{tabular}{@{}cccc@{}}
		\toprule
		$z \backslash y$ & $y=0$ & $y=1$ & Sum	\\ \midrule
		$z=0$ & $10.36\%$ & $2.07\%$ & $12.43\%$\\
		$z=1$ & $30.37\%$ & $57.20\%$ & $87.57\%$\\ \midrule
		Sum & $40.73\%$ & $59.27\%$ \\
		\bottomrule
	\end{tabular}
\end{table}
\section{Discussion and Extension}
\subsection{Imperfect Specificity of the Established Test and the Use of Additional Data} 
\autoref{assumption:no_false-pos} might be too strong for some applications. Although, this assumption simplifies the algebraic expression sometimes significantly, it is not a crucial assumption conceptually. The crucial characterization of joint distributions in \autoref{section:characterization_xyz} can easily be extended to allow for false-positives of the established test.

Evaluations of a new test sometimes have more data available than just $P(y,z|t=1)$. For example, blood samples from before the existence of a virus can serve as true-negative samples. On the other hand, specific blood samples could be analyzed with more sophisticated (and usually much more expensive) methods than just using an established test as reference. These methods would lead to samples with true positives (or at least with very high probability).\footnote{For example, \cite{olbrich-etal-2020} combine these methods to evaluate SARS-CoV-2 antibody tests.} Either of these methods would be provide additional data and therefore would also reduce the missing data problem. In general, this supplementary knowledge leads to narrower bounds, but unless these extra methods are performed for the whole tested population, the missing correlation issues remains. Of course, these methods cannot be applied for the untested population. Therefore the the missing data on prevalence cannot be avoided with these extraneous data.

\subsection{Beyond Binary Tests} \label{subsection:addtionaldata}
The current framework only allows for binary outcomes for both tests and also for the underlying health state. This seems to be the most common situation studied in the literature on diagnostic testing. \citep{zhou-etal-2014} Often tests provide ternary results (with the additional result of `inconclusive' or `invalid'), or allow for even more detailed information, like the Cycle Threshold Count of a PCR test as mentioned in \autoref{subsection: BiN}. In such situations, the theoretic analysis does not provide the appropriate machinery. However, the crucial application to characterize the set of all joint distribution is a result in copula theory \citep[Theorem 3.10]{joe-1997}, which does not rely on any dimension being binary. Indeed, the result even works for continuous outcomes on each dimension.

Similarly, one could use other results in \citet{joe-1997} to characterize the set of possible joint distributions if multiple tests are conducted simultaneously as studied in \citet[Chapter 9]{zhou-etal-2014}. In this case, and like in the characterization of \autoref{section:characterization_xyz}, the testing data are higher-dimensional marginal distribution of an overall joint distribution with an additional dimension (the health state). When considering such an extension, a caution has to be taken because sometimes sharp bounds on the set of possible higher-dimensional distributions may not be known.

\subsection{COVID-19 related testing}
Since testing has a potentially big impact on the economy, an accurate description of the available testing technology is crucial. From the microeconomic perspective, the testing technology affects how test should be optimally allocated (see \cite{ely-etal-2020}, \cite{lipnowski-ravid-2020}) and also how much people engage in social distancing as studied by \cite{acemoglu-etal-2020}. But also the macroeconomy is highly affected by testing strategies and an optimal choice might reduce the economic costs of pandemics considerably. \citep{alvarez-etal-2020,eichenbaum-etal-2020} Although, these studies establish the importance of testing and also address varying testing technologies, all of them assume that a test corresponds to an experiment \`{a} la \cite{blackwell-1951} and therefore is always informative (sometimes the assumption is even that the test itself provides perfect information). 

This paper demonstrates that the assumption of unambiguous information in test results is only applicable if a perfect reference is available when evaluating new tests. In particular, new Antigen test for detection of SARS-CoV-2 are evaluated relative to an imperfect PCR test and therefore---as shown in this paper---these Antigen test produce ambiguous information. An optimal testing procedure should take this ambiguity into account. Similarly, practitioner guides (like \cite{galeotti-etal-2020,watson-etal-2020}) might want to consider addressing uncertainty in test results in more detail.

\newpage
\appendix

\appendix
\section[Characterization of the set of joint distributions]{Characterization of the set of joint distributions $P(x,y,z)$} \label{section:characterization_xyz}
Consider a fixed $\Gamma := P(y=1) \in [\tau \gamma, \gamma]$ with corresponding prevalence $p:=P(x=1) =\Gamma/\sigma$. 
\begin{lemma} \label{lemma:pxy}
	Suppose \autoref{assumption:non-trivial} and \autoref{assumption:no_false-pos} hold. For a given $\Gamma$, $P(x,y)$ is given by \autoref{tab:distribution_pxy}.
\begin{table}[ht!]
	\centering
	\caption{Joint distribution of $P(x,y)$}
	\label{tab:distribution_pxy}
	\begin{tabular}{@{}cccc@{}}
		\toprule
		$P(x \backslash y)$ & $y=0$ & $y=1$ & 	\\ \midrule
		$x=0$ & $1-\nicefrac{\Gamma}{\sigma}$ & $0$ & $1-\nicefrac{\Gamma}{\sigma}$\\
		$x=1$ & $\Gamma \frac{1-\sigma}{\sigma}$ & $\Gamma$ & $\nicefrac{\Gamma}{\sigma}$\\ \midrule
		& $1-\Gamma$ & $\Gamma$ \\
		\bottomrule
	\end{tabular}
\end{table}
\end{lemma}
\begin{proof}
	\begin{enumerate}
		\item $P(x=1,y=1)=P(y=1) - \underbrace{P(x=0, y=1)}_{= 0 \text{ by \autoref{assumption:no_false-pos}}} = \Gamma$
		\item $P(x=1,y=0)=P(x=1) - P(x=1, y=1)=\Gamma(1-\sigma)/\sigma$.
		\item $P(x=0,y=0)=P(y=0)- P(x=1,y=0)= 1 - \Gamma/\sigma$.
	\end{enumerate}	
\end{proof}

Using the law of total probability and rearranging gives
$
	P(y=1|t=0) = \frac{\Gamma-\tau\gamma}{1-\tau}
$.
Furthermore, by the nature of testing $P(y=0,z=1|t=0)=P(y=1,z=1|t=0) =0$. Therefore, $P(y,z)$ is given by \autoref{tab:distribution_pyz}.
\begin{table}[ht!]
	\centering
	\caption{Joint distribution of $P(y,z)$}
	\label{tab:distribution_pyz}
	\begin{tabular}{@{}cccc@{}}
		\toprule
		$P(z \backslash y)$ & $y=0$ & $y=1$ & 	\\ \midrule
		$z=0$ & $1-\Gamma-P(y=0, z=1|t=1)\tau$ & $\Gamma-P(y=1, z=1|t=1)\tau$ & $1-\tau\zeta$\\
		$z=1$ & $P(y=0, z=1|t=1)\tau$ & $P(y=1, z=1|t=1)\tau$ & $\tau\zeta$\\ \midrule
		& $1-\Gamma$ & $\Gamma$ \\
		\bottomrule
	\end{tabular}
\end{table}

By \citet[Theorem 3.10]{joe-1997} the set of all distributions $P(x,y,z)$ with marginals given by $P(x,y)$ and $P(y,z)$ is bounded by two extreme distributions:\footnote{The set of all these distributions, often called Fr\'{e}chet class, is a convex set. Hence, here it suffices to consider the extreme points only.} $\underline{F}_\Gamma \leq F  \leq \overline{F}_\Gamma$, where $F$ is the CDF corresponding to $P(x,y,z)$, $\underline{F}_\Gamma$ is given by \autoref{tab:triple_cdf_lower_Gamma}, and  $\overline{F}_\Gamma$ is given by \autoref{tab:triple_cdf_upper_Gamma}.
\begin{table}[ht!]
	\centering
	\caption{CDF $\underline{F}_\Gamma$}
	\label{tab:triple_cdf_lower_Gamma}
	\resizebox{\textwidth}{!}{
	\begin{tabular}{@{}rcclcc@{}}
		\toprule
		& \multicolumn{2}{c}{$x=1$} &  & \multicolumn{2}{c}{$x=0$} \\  \cline{2-3} \cline{5-6}
		& $z=1$		& $z=0$			&  & $z=1$      			& $z=0$ 													\\ \midrule
		$y=1$ 	& $1$        & $1-\tau \zeta$  		&  & $1-\frac{\Gamma}{\sigma}$	& $\max\left\{0, 1-\nicefrac{\Gamma}{\sigma}- P(y=0, z=1|t=1)\tau\right\}$\\[0.5ex]
		$y=0$ 	& $1-\Gamma$   & $1-\Gamma-P(y=0, z=1|t=1)\tau$	&  & $1-\frac{\Gamma}{\sigma}$	& $\max\left\{0, 1-\nicefrac{\Gamma}{\sigma}- P(y=0, z=1|t=1)\tau\right\}$	\\ \bottomrule
	\end{tabular}}
\end{table}
\begin{table}[ht!] 
	\centering
	\caption{CDF $\overline{F}_\Gamma$}
	\label{tab:triple_cdf_upper_Gamma}
	\resizebox{\textwidth}{!}{
	\begin{tabular}{@{}rcclcc@{}}
		\toprule
		& \multicolumn{2}{c}{$x=1$} 	&  & \multicolumn{2}{c}{$x=0$} 														\\ \cline{2-3} \cline{5-6}
		& $z=1$      & $z=0$      		&  & $z=1$      			& $z=0$ 												\\ \midrule
		$y=1$ 	& $1$        & $1-\tau \zeta$      	&  & $1-\frac{\Gamma}{\sigma}$	& $\min\left\{1-\frac{\Gamma}{\sigma}, 1-\Gamma-P(y=0, z=1|t=1)\tau\right\}$	\\[0.5ex]
		$y=0$ 	& $1- \Gamma$   & $ 1-\Gamma-P(y=0, z=1|t=1)\tau$	&  & $1-\frac{\Gamma}{\sigma}$	& $\min\left\{1-\frac{\Gamma}{\sigma}, 1-\Gamma-P(y=0, z=1|t=1)\tau\right\}$	\\ \bottomrule
	\end{tabular}}
\end{table}

Since $\underline{F}_\Gamma$ and $\overline{F}_\Gamma$ are both nonincreasing in $\Gamma$, sharp bounds for the CDF $F$ across all $\Gamma := P(y=1) \in [\tau \gamma, \gamma]$ are $\underline{F}:=\underline{F}_{\gamma} \leq F  \leq \overline{F}_{\tau\gamma}=:\overline{F}$. For the lower, we have $1-\nicefrac{\Gamma}{\sigma}- P(y=0, z=1|t=1) = (1-\gamma)(1-\tau)+\tau $.

For the upper bound, note that $1-\tau\gamma-P(y=0, z=1|t=1)\tau = 1 - \tau(1-P(y=0, z=0|t=1))$. The corresponding probability mass functions are given by \autoref{tab:triple_pmf_lower} and \autoref{tab:triple_pmf_upper}.
\begin{table}[ht!]
	\centering
	\caption{Lower bound PMF with $\mathcal{P}_{01} := P(y=0, z=1|t=1)$}
	\label{tab:triple_pmf_lower}
	\resizebox{\textwidth}{!}{
		\begin{tabular}{@{}rcclcc@{}}
			\toprule
			& \multicolumn{2}{c}{$x=1$} &  & \multicolumn{2}{c}{$x=0$} \\  \cline{2-3} \cline{5-6}
			& $z=1$		& $z=0$			&  & $z=1$      			& $z=0$ 													\\\midrule
			$y=1$ 	& $P(y=1, z=1|t=1)\tau$	& $\gamma-P(y=1, z=1|t=1)\tau$  	&  & $0$	& $0$	\\[1ex]
			$y=0$ 	& $\max\left\{0,  \frac{\gamma}{\sigma}+ \mathcal{P}_{01}\tau-1\right\}$	& $\min\left\{\gamma \frac{1-\sigma}{\sigma},1 - \gamma - \mathcal{P}_{01}\tau\right\}$	&  & $\min\left\{1-\frac{\gamma}{\sigma}, \mathcal{P}_{01}\tau\right\}$	& $\max\left\{0, 1-\frac{\gamma}{\sigma}-\mathcal{P}_{01}\tau\right\}$	\\[1ex] \bottomrule
	\end{tabular}}
\end{table}

\begin{table}[ht!]
	\centering 
	\caption{Upper bound PMF with $\mathcal{P}_{00} := P(y=0, z=0|t=1)$}
	\label{tab:triple_pmf_upper}
	\resizebox{\textwidth}{!}{
		\begin{tabular}{@{}rcclcc@{}}
			\toprule
			& \multicolumn{2}{c}{$x=1$} 																								&  & \multicolumn{2}{c}{$x=0$} 														\\ \cline{2-3} \cline{5-6}
			& $z=1$      													& $z=0$      												&  &
			$z=1$      														& $z=0$ 												\\ \midrule
			$y=1$ 	& $P(y=1, z=1|t=1)\tau$									& $P(y=1, z=0|t=1)\tau$      										&  &
			$0$																& $0$	\\ [1ex]
			$y=0$ 	& $\tau\left[\min\left\{\frac{\gamma}{\sigma}, 1-\mathcal{P}_{00}\right\} -\gamma\right]$	& $\tau\max\left\{\mathcal{P}_{00}+\frac{\gamma}{\sigma}-1,0 \right\}$	&  &
			$\tau\max\left\{1-\frac{\gamma}{\sigma}-\mathcal{P}_{00},0 \right\}$			& $1-\tau\max\left\{\frac{\gamma}{\sigma}, 1-\mathcal{P}_{00}\right\}$	\\[1ex] \bottomrule
	\end{tabular}}
\end{table}

\bibliographystyle{ecta} 
\bibliography{dilation_tests}


\end{document}